\documentclass[journal=jpclcd,manuscript=article]{achemso}

\usepackage[version=3]{mhchem} 

\author{Sandro Bottaro}
\affiliation{SISSA, International School for Advanced Studies 265, Via Bonomea I-34136 Trieste, Italy}
\email{sbottaro@sissa.it}

\author{Pavel Ban\'a\v{s}}
\affiliation{Regional Centre of Advanced Technologies and Materials, Department of Physical Chemistry, Faculty of Science, Palacky University Olomouc, 17. Listopadu 12, 771 46 Olomouc, Czech Republic}
\alsoaffiliation{Institute of Biophysics, Academy of Sciences of the Czech Republic, Kralovopolska 135, 612 65 Brno, Czech Republic}            
\author{Ji\v{r}\'i \v{S}poner}
\affiliation{Regional Centre of Advanced Technologies and Materials, Department of Physical Chemistry, Faculty of Science, Palacky University Olomouc, 17. Listopadu 12, 771 46 Olomouc, Czech Republic}
\alsoaffiliation{Institute of Biophysics, Academy of Sciences of the Czech Republic, Kralovopolska 135, 612 65 Brno, Czech Republic}            

\author{Giovanni Bussi}
       \affiliation{SISSA, International School for Advanced Studies 265, Via Bonomea I-34136 Trieste, Italy}
\email{bussi@sissa.it}

\title{Free Energy Landscape of GAGA and UUCG RNA Tetraloops}

\begin{document}
\vspace{1cm}

\begin{tocentry}
\includegraphics{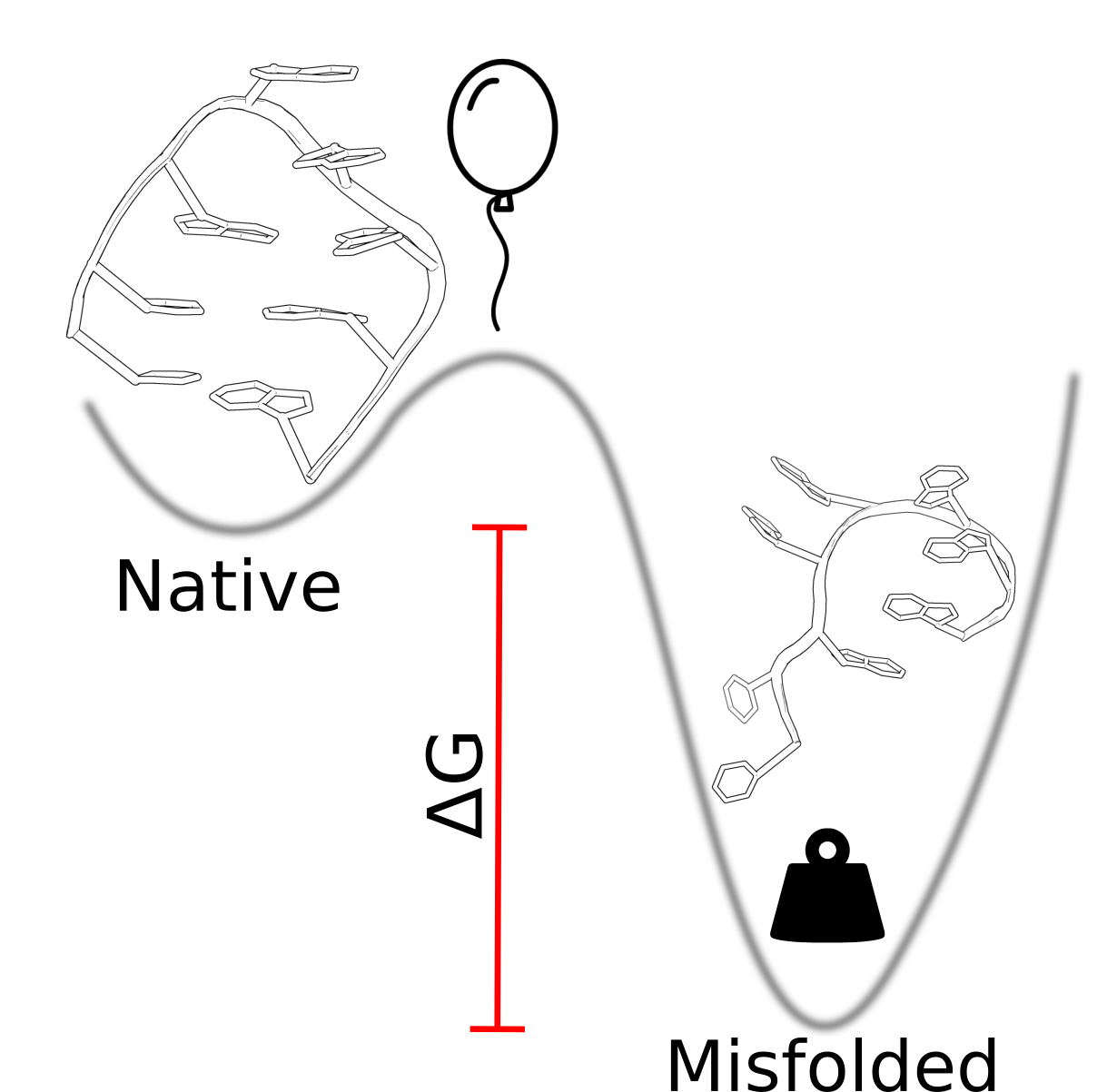}
\end{tocentry}
Reprinted with permission from:\\
``Free Energy Landscape of GAGA and UUCG RNA Tetraloops''\\
Authors: Sandro Bottaro, Pavel Banas, Jiri Sponer, and Giovanni Bussi.\\
Journal: Journal of Physical Chemistry Letters.\\
doi: 10.1021/acs.jpclett.6b01905. Copyright 2016 American Chemical Society.\\

\begin{abstract}
  We report the folding thermodynamics of ccUUCGgg and ccGAGAgg RNA tetraloops using atomistic molecular dynamics simulations. 
We obtain a previously unreported estimation of the folding free energy using parallel tempering in combination with well-tempered metadynamics. A key ingredient is the use of a recently developed metric distance, eRMSD, as a biased collective variable. We find that the native fold of both tetraloops is not the global free energy minimum using the Amber$\chi_{\text{OL3}}$ force field. The estimated folding free energies are 30.2$\pm$0.5 kJ/mol for UUCG and 7.5$\pm$0.6 kJ/mol for GAGA, in striking disagreement with experimental data.   
  We evaluate the viability of all possible one-dimensional backbone force field corrections. We find that disfavoring the \textit{gauche}$^+$ region of $\alpha$ and $\zeta$ angles consistently improves the existing force field. The level of accuracy achieved with these corrections, however, cannot be considered sufficient by judging on the basis of available thermodynamic data
and solution experiments. 
\end{abstract}


RNA tetraloops are small, stable and ubiquitous three-dimensional motifs playing an important structural and functional role in many RNA molecules~\cite{woese1990architecture}. Tetraloops consist of an A-form helical stem capped by 4 nucleotides arranged in a very specific three-dimensional structure. The  great majority of known RNA tetraloops structures have sequence GNRA or UNCG, where N is any nucleotide and R is guanine or adenine. Their small size, together with the abundance of experimental data available, made these systems primary targets for atomistic molecular dynamics (MD) simulation studies~\cite{Zichi1995}. While simulations initialized in the vicinity of the native state are stable on short time-scales under a variety of simulation conditions~\cite{Miller1997,Williams1999,Sorin2002,Villa2006,Ferner2008,Giambasu2015}, more recent works strongly suggest that these systems are not correctly modeled by the current Amber force field~\cite{banas2010performance,Kuhrova2013,Haldar2015,bergonzo2015highly,Kuhrova2016,hase2016free}. Although different improvements have been proposed,~\cite{Chen2013} there is growing evidence that none of the available corrections are able to capture the crucial non-canonical interactions present in these tetraloops~\cite{bergonzo2015highly,Kuhrova2016}.
Despite their small size, an ergodic sampling of these systems requires substantial computational resources, in the order of hundreds of $\mu$s using massively parallel simulations~\cite{Sorin2002,Kuhrova2013,bergonzo2015highly}. For this reason, full convergence of MD simulations on RNA has been so far achieved for simple systems such as tetranucleotides~\cite{bergonzo2013multidimensional}.

In this paper we address three different problems connected to MD simulations of RNA tetraloops. First, we show that an accurate estimation of the folding free energy of ccGAGAgg and ccUUCGgg tetraloops can be achieved by combining parallel tempering with well-tempered metadynamics. This combination enhances the efficiency of parallel tempering alone, thus greatly reducing the computational cost of these simulations. A fundamental aspect of our work is the use of the eRMSD~\cite{bottaro2014role} as a biasing collective variable. The eRMSD is a metric for measuring distances between RNA structures which has been recently suggested as a non-trivial replacement of the common but inadequate RMSD.  The eRMSD is based on the relative position and orientation of nucleobases only, and it has been shown to precisely and unambiguously discriminate between different RNA structures. It is very accurate and effective in reflecting structural and dynamical features of RNA molecules as well~\cite{bottaro2016rna,Kuhrova2016}. 
Second, we provide a definition of the native ensemble that is compatible with available solution Nucleic Magnetic Resonance (NMR) data. Lastly, we evaluate viable backbone force field correction terms, with the aim of improving the current Amber force field. We find that improved accuracy is achieved by disfavoring the \textit{gauche}$^+$ region of $\alpha$ and $\zeta$ torsion angles as it was recently proposed~\cite{gil2016empirical}. These corrections highly stabilize the native fold of GAGA as well as of five different RNA tetranucleotides, while a modest improvement is observed for the UUCG tetraloop. The simulation protocol introduced here is robust and makes it possible to unambiguously assess a force-field's accuracy and to quantitatively evaluate the impact of force field modifications.

Molecular dynamics simulations were performed using the GROMACS 4.6.7 software package~\cite{pronk2013gromacs} in combination with PLUMED 2.2~\cite{tribello2014plumed}.
Ideal A-form, fully stacked initial conformations were generated using the Make-NA web server (http://structure.usc.edu/make-na/server.html).
The systems described in Table 1 were solvated in a truncated dodecahedral box with TIP3P~\cite{jorgensen1983comparison} water molecules and neutralized by adding Na$^+$ counterions~\cite{joung2008determination}. RNA was modeled using the Amber99 force field\cite{cornell1995second} with parmbsc0\cite{perez2007refinement} and $\chi_{\text{OL3}}$ corrections \cite{banas2010performance}. We will refer to this combination of force field corrections as Amber$\chi_{\text{OL3}}$ in the remainder of this paper. Parameters are available at http://github.com/srnas/ff.
The initial conformations were minimized in vacuum first, followed by a minimization in water and equilibration in NPT ensemble at 300K and 1 bar for 1 ns.
Production runs were performed in the canonical ensemble using stochastic velocity rescaling thermostat~\cite{bussi2007canonical}. All bonds were constrained using the LINCS algorithm~\cite{hess1997lincs}, equations of motion were integrated with a time step of 2 fs. 
Temperature replica exchange MD~\cite{sugita1999replica} in combination with well-tempered metadynamics~\cite{laio2002escaping,barducci2008well} was used to accelerate sampling~\cite{bussi2006free}. For each system 24 geometrically-distributed replicas in the temperature range 278K-400K were simulated for 1.0 $\mu$s per replica. The average acceptance rate varied from $\approx$2\% to 9\%, depending on the system and on the temperature, with an average round-trip time between 4 and 9 ns (see Table 1). The eRMSD~\cite{bottaro2014role} from the native reference structure was used as a biasing collective variable. A short description of the eRMSD metric, and a comparison with the standard RMSD measure is presented in Figure S1. The native reference structure for GAGA tetraloop was taken from the crystal structure of the SAM-I/IV riboswitch (PDB code 4L81, residues 75-82)~\cite{trausch2014structural}, while the UUCG native structure was taken from the PDB structure 1F7Y, residues 7-14. 
 A Gaussian bias of width 0.1 was deposited every 500 steps, and the initial height of 0.5 kJ/mol was decreased with a biasfactor 15. The implementation of the eRMSD as a collective variable is available on request and will be included in PLUMED 2.3. A sample PLUMED input file can be found in Supporting Text S1. Samples from simulations were analyzed every ps and excluding the first 200ns. After the first 200ns the fluctuations of the bias potential are negligible due to the damping of Gaussian height. 
Statistical errors were estimated using blocks of 200ns. Free energy surfaces  obtained from such blocks are substantially identical, leading to a statistical error below k$_{\text B}$T.  We notice that metadynamics induces an almost uniform sampling on the biased collective variable, minimizing its statistical error. However, the statistical error associated to the population of conformers that are not necessarily distinguished by eRMSD from native, such as the end-to-end distance, could be larger.
 Free energy differences were calculated as $\Delta G = -k_\text{B}T [ \log(\sum_{i,\text{folded}}p(x_i)) - \log(\sum_{i,\text{unfolded}}p(x_i)) ]$. eRMSD thresholds for the folded and unfolded regions, chosen based on the peaks of the free energy surface, are listed in Table 1. MD simulations of tetranucleotides were taken from previous studies~\cite{gil2016empirical,bottaro2016rna}. Reweighting was performed using the final bias potential.~\cite{branduardi2012metadynamics}

\begin{table}[ht]
\centering
\begin{footnotesize}
\begin{tabular}{c c c c c c}
System & sequence & N. Water & eRMSD fold$^{a}$ & Acc. rate (\%)$^b$ & $\tau$(ns)$^c$\\ 
GAGA 6 & cGAGAg    & 3000  & 0.7 & 3.7-7.3 & 5.3\\
GAGA 8 & ccGAGAgg  & 3766  & 0.8 & 2.0-4.6 & 8.6 \\
UUCG 6 & cUUCGg    & 2719  & 0.7  & 4.5-8.6 & 4.1 \\
UUCG 8 & ccUUCGgg  & 3915  & 0.72 & 1.8-4.7 & 8.6\\
\end{tabular}
\end{footnotesize}
\caption{Simulation details.(a) eRMSD threshold used for calculating free energy differences between folded and unfolded. (b) Average acceptance rate for the cold (278K) and hot (400K) replicas. (c) Average roundtrip time for all replicas.}
\end{table}

\begin{figure}
\centering
\includegraphics{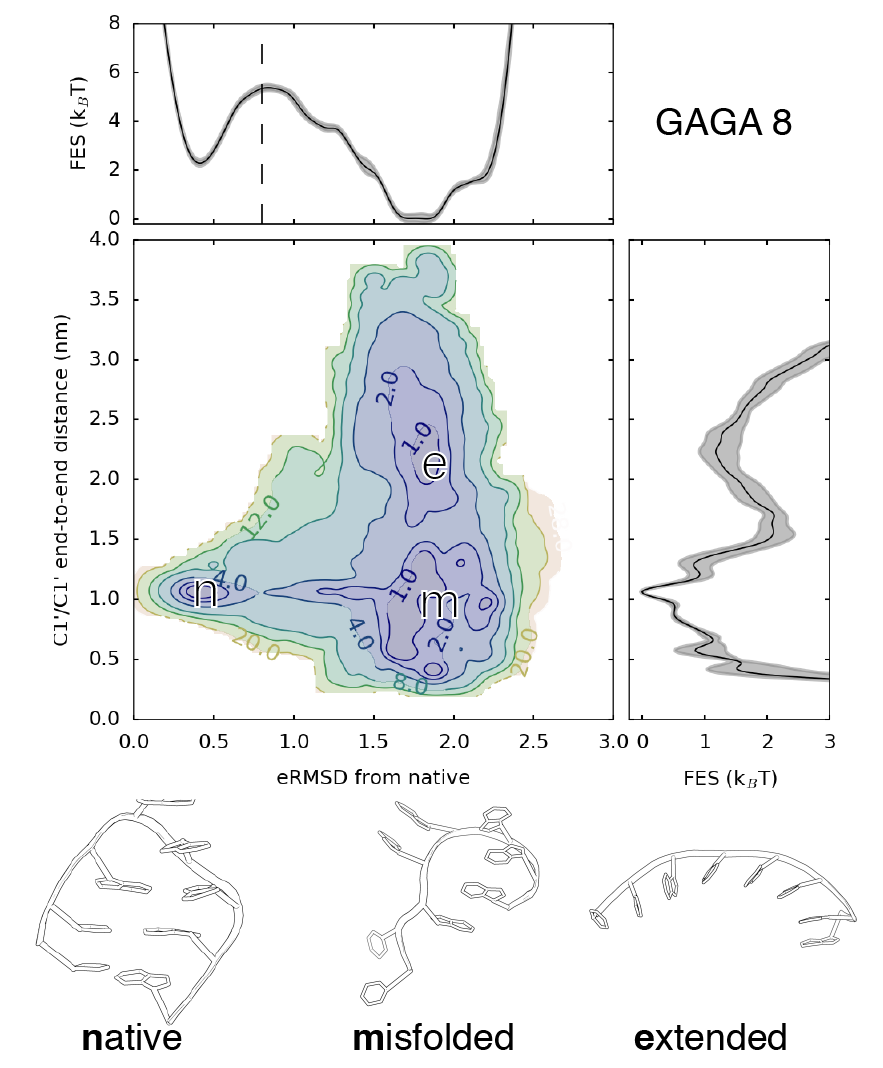}
\caption{Free energy surfaces projected onto the eRMSD from native and onto the end-to-end distance at 300.9K for GAGA tetraloop. In the two-dimensional projection the colors indicate the free energy difference with respect to the minimum. Labels of the isolines are expressed in k$_{\text{B}}$T. Gray shades indicate statistical error. Representative three-dimensional structures for each region discussed in the main text are shown.}
\label{fig:fig1a}
\end{figure}

\begin{figure}
\centering
\includegraphics{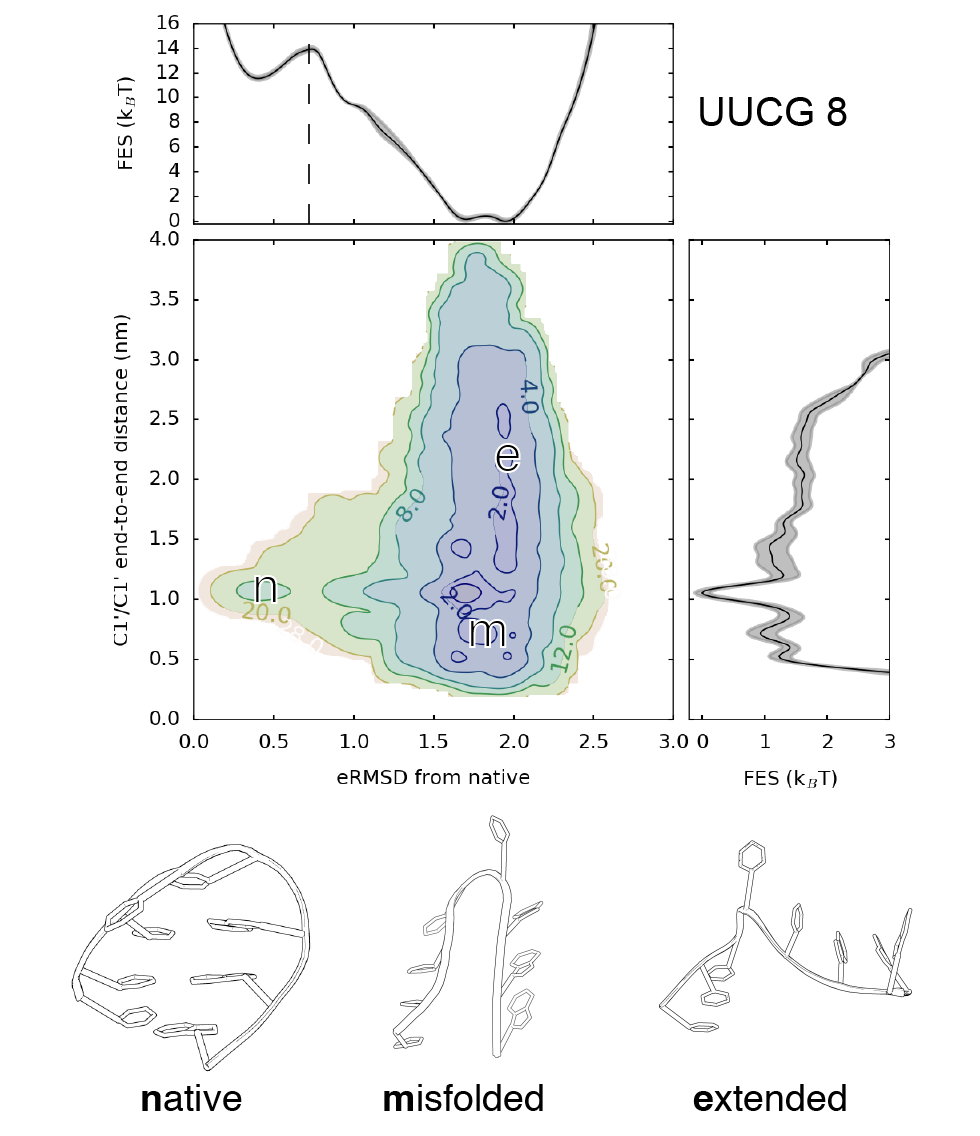}
\caption{Free energy surfaces projected onto the eRMSD from native and onto the end-to-end distance at 300.9K for UUCG tetraloop. In the two-dimensional projection the colors indicate the free energy difference with respect to the minimum. Labels of the isolines are expressed in k$_{\text{B}}$T.  Gray shades indicate statistical error. Representative three-dimensional structures for each region discussed in the main text are shown.}
\label{fig:fig1b}
\end{figure}

\textit{Free energy landscape of GAGA and UUCG tetraloops.} In each WT-REMD simulation multiple folding and unfolding events are observed, as shown in Figure S2.
The free energy surfaces of the GAGA and UUCG tetraloops, projected onto the eRMSD from native and onto the end-to-end distance at 300.9K,  are shown in Fig.\ \ref{fig:fig1a} and \ref{fig:fig1b}.

For the GAGA tetraloop, the global free energy minimum consists of misfolded, compact conformations characterized by several non-native stacking interactions and by the absence of the two Watson-Crick base pairs in the stem. Similar highly stacked structures were reported in previous simulation studies~\cite{condon2015stacking,bergonzo2015highly,bottaro2016rna}. We also observe a second local minimum composed by extended structures (end-to-end distance$>$1.5 nm).
The native basin (eRMSD$<$0.8) is considerably less stable compared to unfolded/misfolded conformations by 7.5$\pm$ 0.6 kJ/mol.
Note that this result is not compatible with the prediction of $\Delta$G=-2.8 kJ/mol obtained using the nearest neighbor parameters~\cite{hofacker2003vienna,mathews2004incorporating}.
Here, we employed a rather strict definition of the native basin. Qualitatively, the threshold of eRMSD$<$0.8 corresponds to fully native structures where the stem is correctly formed and the trans sugar/Hoogsteen (tSH) G1-A4 non-canonical base-pair is present, while only the apical base A2 is flexible. Although it is not possible to establish a one-to-one mapping between eRMSD and standard RMSD measure, all structures with eRMSD$<$0.8 from native have an RMSD distance lower than 3\AA\ (see also Fig.\ S1).    
We notice that the flexibility of A2 has been also reported in a recent analysis of crystallographic structures~\cite{bottaro2016rna}.
Similar results ($\Delta$G=6.8$\pm$0.6 kJ/mol) are obtained using a broader definition of the native basin, in which only the stem is required to be correctly formed, with no restriction on the loop.
Both these definitions rely on the assumption of a reference structure which corresponds to
the conformation observed in crystallographic databases. In principle, the solution
structure or, better, NMR primary data should be used as a benchmark.
In this specific case, however, available Nuclear Overhauser Effect (NOE)~\cite{jucker1996network} distances are very sparse, making it difficult to carry out such analysis (see Figure S3).

Similarly to GAGA, for the UUCG tetraloop the global free energy minimum is characterized by a short end-to-end distance where none of the native UUCG interactions are present, neither in the stem nor in the loop (misfolded basin in Fig.\ \ref{fig:fig1b}). The conformational ensemble of the global minimum is composed by compact structures with different stacking arrangements between non-consecutive nucleobases and stabilized by additional base-phosphate hydrogen bonds.
Compatibly with the presence of several consecutive pyrimidine bases, the local minimum corresponding to extended conformations is less pronounced compared to the GAGA tetraloop. The native basin, here defined based on the peak of the barrier at eRMSD=0.72, has a free energy difference with respect to the unfolded basin of 30.2$\pm$0.5 kJ/mol, considerably higher compared to the nearest neighbor prediction of $\Delta$G=-1.97 kJ/mol. NOE distances unambiguously support the use of a strict definition of the native basin that takes into account both stem and loop (see Figure S4). When requiring only the stem to be formed to consider the tetraloop as folded, we obtain a folding free energy difference of 15.0$\pm$2.6 kJ/mol. However, such definition of the folded state is not supported by the abundant NOE distances available for this system~\cite{nozinovic2010high}.  
The free energy surfaces of GAGA and UUCG hexamers (Figure S5) are in agreement with the findings reported above, further confirming that over-hydrogen-bonded, compact and non-native conformations are highly overstabilized.

\begin{figure}
\centering
\includegraphics[scale=0.9]{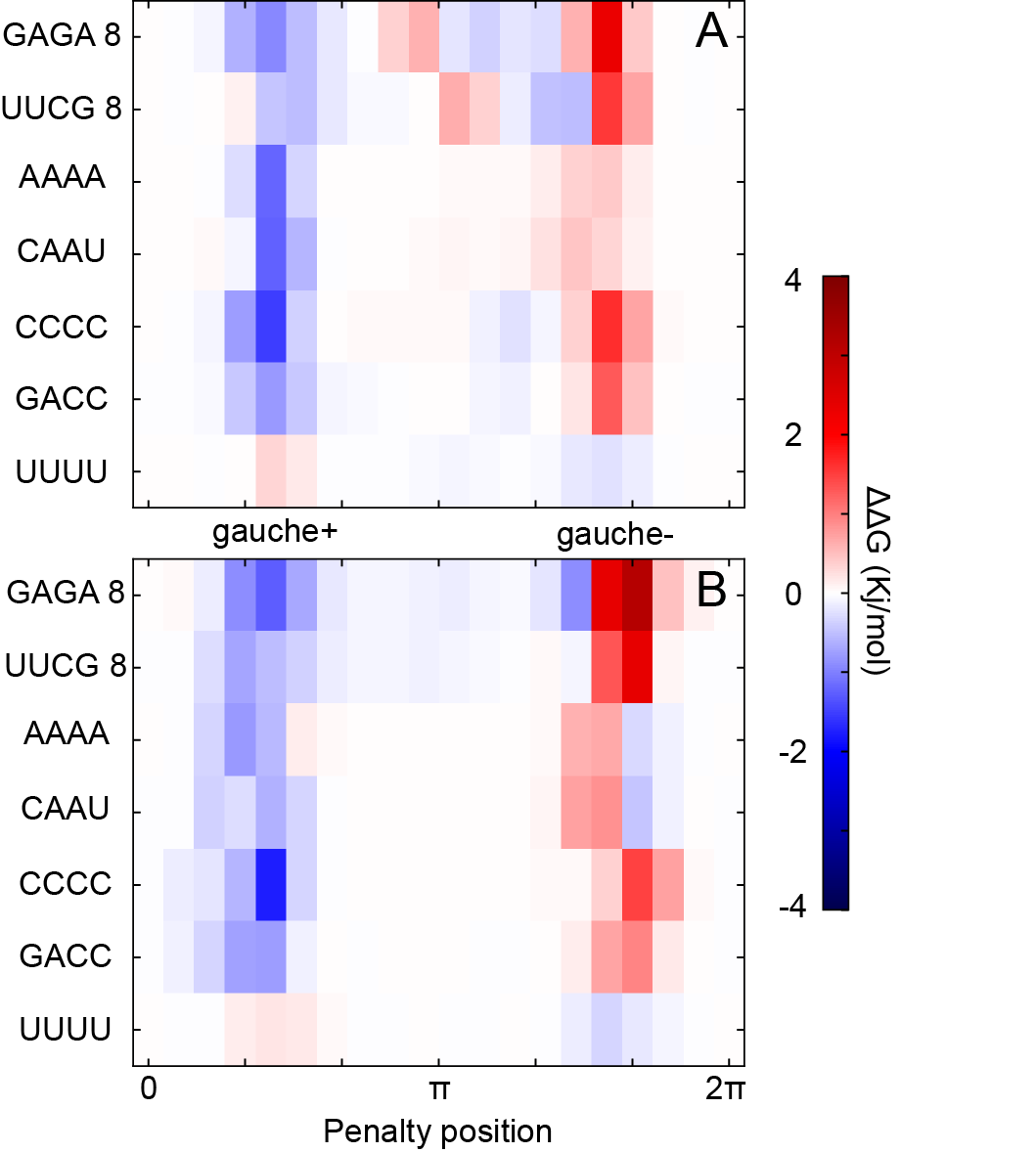}
\caption{Effect on the stability of the native fold (for UUCG/GAGA) and of extended conformation upon addition of a Gaussian potential on $\alpha$ (panel A) and $\zeta$ angles (panel B). Blue indicates that the additional potential stabilizes the correct structure, red indicates that the native fold is destabilized.}
\label{fig:fig2b}
\end{figure}
\textit{Exploring the impacts of torsional corrections.}
The Amber$\chi_{\text{OL3}}$ force field for RNA significantly underestimates the stability of the native fold for both GAGA and UUCG tetraloop. We here seek an answer to the following question: is it possible to introduce a torsion backbone correction that systematically improves the stability of the native folds? To this end, we added a local potential disfavoring specific backbone conformation for each torsion angle ($\alpha$,$\beta$,$\gamma$,$\delta$,$\epsilon$,$\zeta$,$\chi$), and calculated the change in the folding free energy upon the addition of the bias. For simplicity, the new folding free energies are calculated using standard reweighting techniques. We obtain an estimate of the functional derivative of the stability with respect to arbitrary modification of the torsional potential by performing a systematic scan using a Gaussian potential with height 2 kJ/mol and sigma 0.13 rad.
In order to ensure the transferability of the corrections, we additionally analyzed MD trajectories on 5 tetranucleotides~\cite{gil2016empirical,bottaro2016rna}. NMR studies showed these tetranucleotides to be mostly in A-form-like conformation in solution~\cite{condon2015stacking,tubbs2013nuclear}. At variance with experimental evidence, previous MD simulations showed that compact, interdigitated structures are over-stabilized by the Amber$\chi_{\text{OL3}}$ force field~\cite{henriksen2013reliable,bergonzo2015highly,bottaro2016rna}. We thus expect viable force field corrections to improve the agreement with experiments on these systems as well.
Figure \ref{fig:fig2b} shows the change in stability of the native fold upon addition of the Gaussian penalty as a function of its position. It can be seen that systematic improvements can be obtained by  penalizing the \textit{gauche}$^+$ region in $\alpha$ and $\zeta$, in agreement with a previous simulation study~\cite{gil2016empirical}.  Corrections to the remaining backbone angles ($\beta$,$\gamma$,$\delta$, $\epsilon$ and $\chi$) have contrasting or not significant impact on the stability. In particular, penalizing high-anti conformer in $\chi$ angles increases the stability of the tetraloops, but it has detrimental effects on tetranucleotides (see Figure S6). It has been observed that penalizing high-anti conformers can also lead to a flattening of the A-helix geometry~\cite{banas2010performance}.

\begin{figure}
\centering
\includegraphics[scale=0.9]{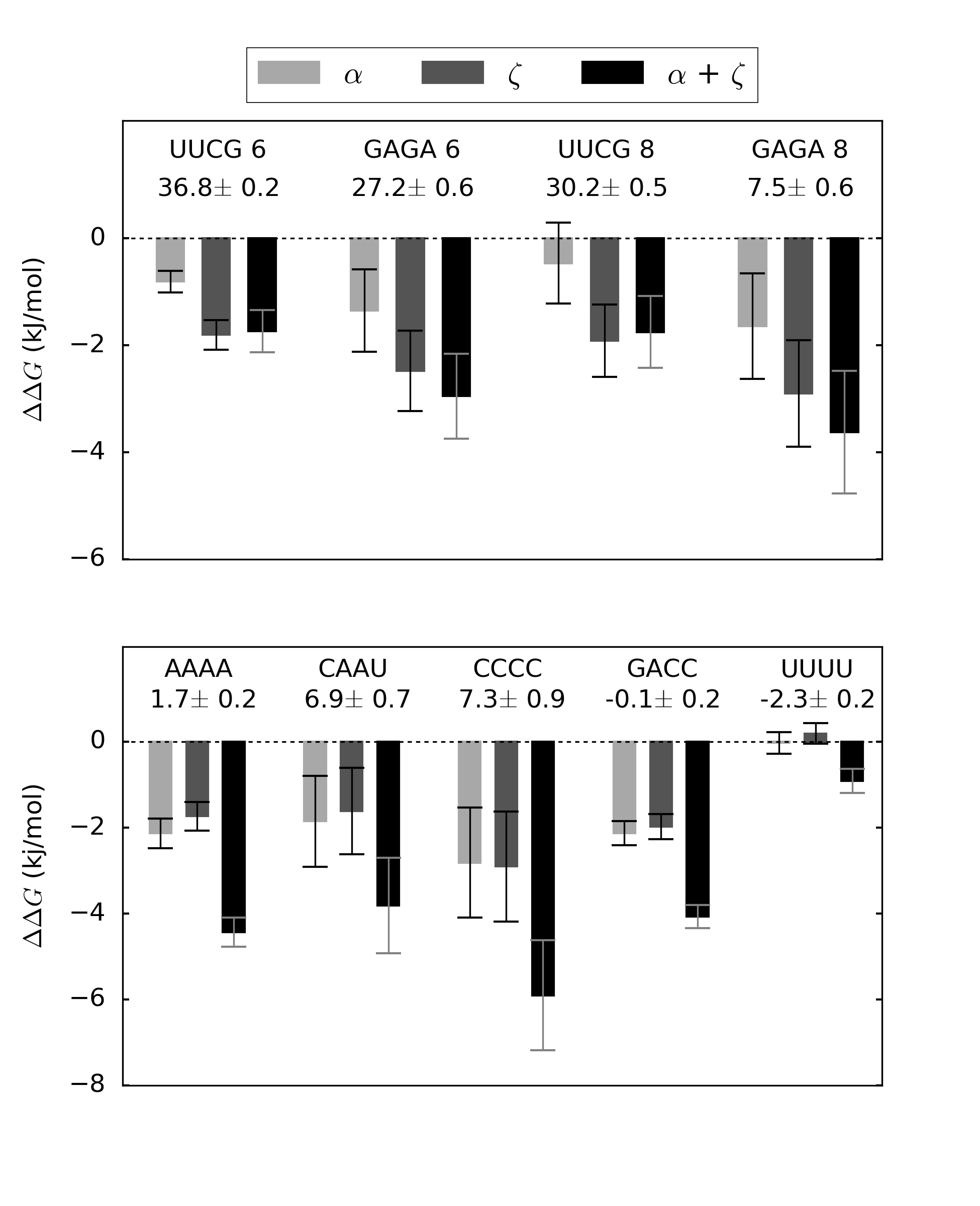}
\caption{Folding free energy changes upon addition of a cosine correction $f(\theta) = \cos(\theta+4.5)$ to $\alpha$, $\zeta$, and $\alpha$+$\zeta$ angles for different tetraloops and tetranucleotides. $\Delta$G (in kJ/mol) calculated using the uncorrected Amber$\chi_{\text{OL3}}$ force field are reported in the figure.}
\label{fig:fig2}
\end{figure}

The results shown in Fig.\ \ref{fig:fig2b} suggest that small adjustments to $\alpha$ and $\zeta$ angles can improve the agreement with experiments of the Amber$\chi_{\text{OL3}}$ force field. The profiles suggest that a cosine with periodicity 2$\pi$ would lead to a consistent improvement. We therefore evaluate the effects of a simple potential correction in the form $f(\theta) = k\cos(\theta+\phi)$. We find that optimal results are obtained by employing a phase $\phi=4.5$rad, while we set $k=1$kJ/mol in order to keep the correction small. Two items are worth highlighting. First, the modification has a minimum in the \textit{gauche}$^-$ region, and as such it is not expected to affect the canonical A-form helix (see Figure S7). Second, this modification can be easily incorporated within a force field, being a standard cosine torsion potential term. In Fig.\ \ref{fig:fig2} we show that both modifications on $\alpha$ and $\zeta$, as well as the combination of the two, lead to a significant stabilization of the native folds in all cases. In particular, the $\alpha$+$\zeta$ modification provides the best results.
For GAGA, the folding free energy drops from 7.5 to 3.9 kJ/mol at 300.9K, using both $\alpha$+$\zeta$ corrections. The UUCG tetraloop is only marginally affected, as the folding free energy diminishes from 30.2 to 28.5 kJ/mol at the same temperature. $\alpha$/$\zeta$ modifications also improve the behavior on all the tetranucleotides, as the over-stacked, over-hydrogen-bonded conformations are destabilized with respect to A-form-like extended structures. Individual free energy surfaces with and without backbone modifications are shown in Figure S8.  

Additionally, we further evaluate the $\alpha$/$\zeta$ corrections with respect to high-level quantum mechanical calculations. We find that our torsion tweaks have small negative effects. More precisely the mean absolute error calculated on a diverse set of UpU dinucleotides~\cite{kruse2015quantum} increases from 6.7 kJ/mol (Amber$\chi_{\text{OL3}}$) to 7.1 kJ/mol ($\alpha$),7.0 ($\zeta$) and 7.6 kJ/mol ($\alpha$+$\zeta$), as shown in Figure S9.
We notice however that this change is likely within the expected accuracy of the reference
quantum mechanical calculations. In this respect, it is worthwhile observing that
small changes in the torsional potential, which are below the accuracy
of the fitting usually done in force field parametrizations, can significantly affect the 
free energy landscape when adopted to simulate larger molecular systems.

\begin{figure}
\centering
\includegraphics{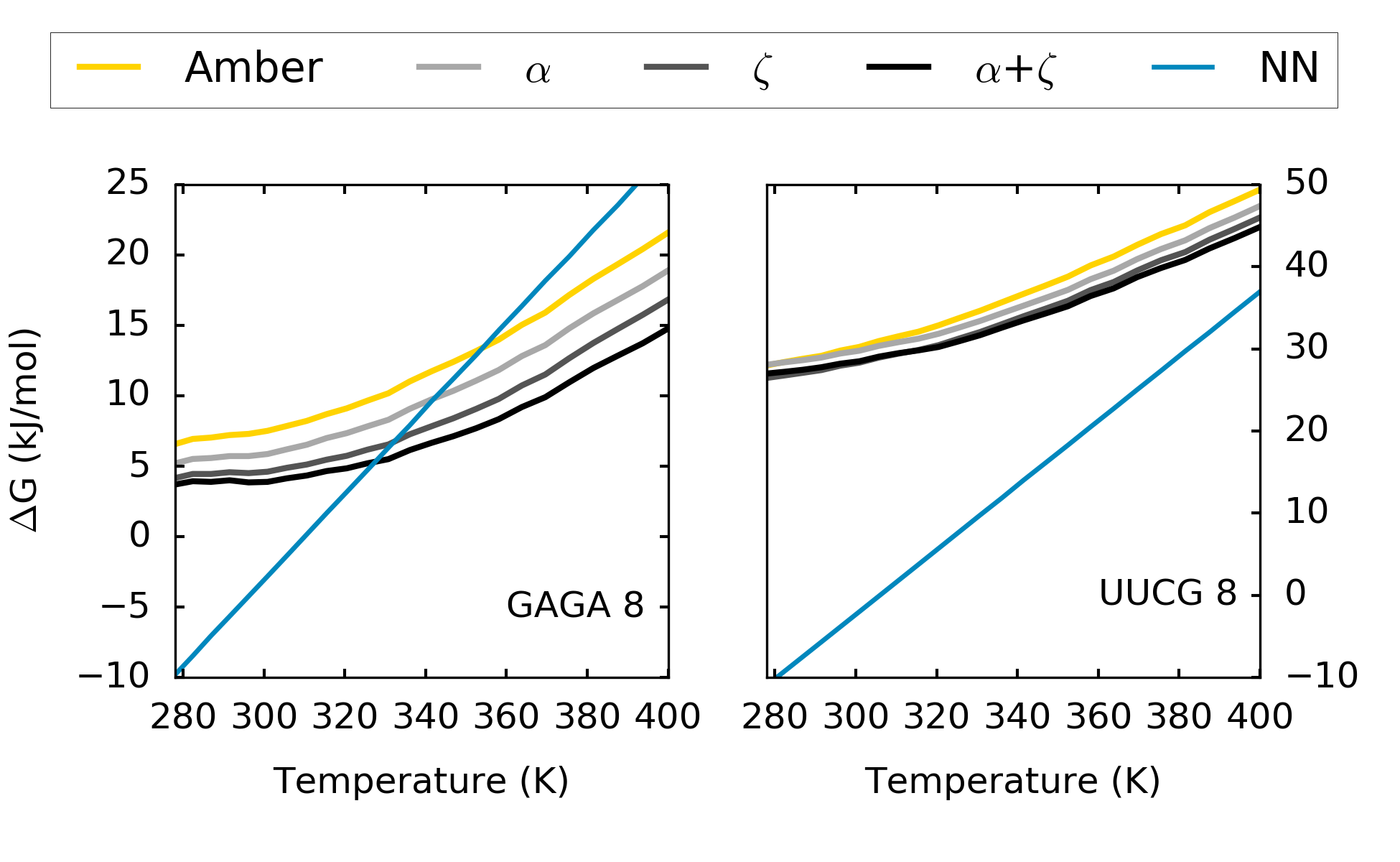}
\caption{Folding free energy as a function of the temperature obtained from MD simulations compared with the prediction obtained from the nearest neighbor (NN) model. Statistical errors, calculated using blocks of 200ns, are in the order of 1kJ/mol and are not shown for clarity. Statistical error is higher in the reweighted ensembles due to the lower effective sample size, but still allows $\Delta G$ to be estimated with a statistical error lower than 2kJ/mol.}
\label{fig:fig3}
\end{figure}

In Fig.\ \ref{fig:fig3} we show the temperature dependence of the folding free energy for the two tetraloops,
compared with the prediction obtained using optical melting experiments data (nearest neighbor model).  The nearest neighbor prediction typically deviates from experimental data by 2-3 kJ/mol on GNRA and UNGC tetraloops~\cite{sheehy2010thermodynamic}. As also described above, the Amber$\chi_{\text{OL3}}$ force field does not reproduce correctly the folding free energy predicted by the nearest neighbor models. The disagreement is less pronounced for GAGA, while it is dramatic for UUCG tetraloop. The backbone corrections are able to consistently shift the MD prediction closer to the experiments, but the level of agreement still remains unsatisfactory. 
While the folding free energy is by construction linearly dependent on the temperature in the nearest neighbor model,
MD simulations display a clear non-linear dependence.
This is different from what has been observed in protein hairpins~\cite{munoz1997folding}
and suggests that a simple two-state model approximation might be not valid for RNA hairpins,
in agreement with previous experimental works~\cite{ma2006exploring}.
We notice however that temperature dependence in replica-exchange simulations performed 
at constant volume could be affected by a spurious high pressure in the high temperature replicas.

Our calculations provide sufficient sampling to reliably derive the folding thermodynamics as predicted by the standard Amber$\chi_{\text{OL3}}$ force field.
This is achieved by using parallel tempering with well-tempered metadynamics.
Whereas parallel tempering increases the ergodicity of the system as a whole,
the metadynamics bias potential flattens the distribution along the
biased collective variable.
As a matter of fact, choosing the eRMSD from native as a biased variable allowed us to
obtain for the first time converged folding free energy landscapes for GAGA and UUCG tetraloops.
Two items here are worth highlighting. First, we found the eRMSD to be a fundamental ingredient in our simulation protocol. In our tests the use of the standard RMSD as biased collective variable
did not allow us to reach the native structure with the correct base-pairing interactions. This result confirms the validity of eRMSD as a structural measure for comparing, clustering, analyzing and modeling RNA structures~\cite{bottaro2014role,bottaro2016rna,Kuhrova2016}.  Second, previous simulation studies using parallel tempering alone on the same systems typically required at least one order of magnitude more computational power.  

 
This study proves that the native folds of GAGA and UUCG tetraloops are not the global minimum of the Amber force-field without any uncertainty due to incomplete sampling, and this study alone is able to quantify by how much they are destabilized compared to experiment. 
The discrepancy with respect to predictions obtained from experimental data is significant for GAGA and critical for UUCG tetraloop. Two effects mainly contribute to this discrepancy: i) overstabilization of highly-stacked, compact structures with no base-pairs~\cite{Kuhrova2016} and ii) under-stabilization of the native pattern of stacking/non-canonical base-pairing.  The difference between the GAGA and UUCG tetraloops indicates that there is either some critical force field deficiency in describing the native interactions in the UUCG tetraloop (e.g. the trans sugar/Hoogsteen base-pair), or that spurious substates in the UUCG misfolded ensemble are significantly overstabilized compared to the GAGA misfolded basin. It is as well plausible that both factors contribute to the highly inaccurate free energy landscape predicted for UUCG.
A careful comparison with solution data performed here indicates that the only definition of folded UUCG tetraloop that is compatible with available NMR data is the one
where all interactions of the consensus crystallographic structure are present. This analysis could not be carried out for the GAGA tetraloop, due to the paucity of available NOE distances.
%

An increasing number of simulation studies pointed out deficiencies of the RNA Amber force fields, and the present study corroborates the fact that substantial improvements are needed in order to use MD simulations in a predictive way, especially for systems with a significant number of non-canonical interactions. 
By reweighting our simulations we introduce backbone corrections to $\alpha$ and $\zeta$ angles that improve the agreement of the simulations with experimental data. 
The level of achieved accuracy does not seem to be sufficient. This suggests that small modifications to the torsion potentials alone would not be able to correct major force-field deficiencies. This is not surprising, since dihedral potentials are formally \emph{intramolecular} energy terms which do not have any corresponding quantum-mechanical observable. They are used for final pragmatic tuning of the force fields, implicitly correcting for errors in a number of real physical terms including \emph{intermolecular} contributions. Thus, the capability of tuning biomolecular force fields by dihedral potentials cannot be unlimited. 
  We propose our simulation protocol as a tool to assess and compare existing force fields for RNA. We additionally 
envisage the possibility of testing recent corrections to the Van der Waals parameters~\cite{steinbrecher2012revised,Chen2013} in conjunction with more accurate water models~\cite{mukhopadhyay2012charge,bergonzo2015improved}.

\acknowledgement
GB and SB have received funding from the European Research Council under the European Union's Seventh Framework Programme (FP/2007-2013) / ERC Grant Agreement n. 306662, S-RNA-S.
JS and PB were supported by grant P208/12/1878 from the Czech Science Foundation. Institutional funding was provided by project LO1305 of the Ministry of Education, Youth and Sports of the Czech Republic (JS and PB). JS acknowledges support by Praemium Academiae.
Alan Chen and Martin Zacharias are acknowledged for carefully reading the manuscript and providing useful suggestions.

\begin{suppinfo}
\textbf{Text S1} Sample PLUMED input file.\\
\textbf{Figure S1} Definition of eRMSD and comparison with RMSD.\\
\textbf{Figure S2} eRMSD from native of demultiplexed trajectories.\\
\textbf{Figure S3} Comparison between experimental and calculated NOE distances for GAGA tetraloop.\\
\textbf{Figure S4} Comparison between experimental and calculated NOE distances for UUCG tetraloop.\\
\textbf{Figure S5} Free energy surfaces projected onto the eRMSD from native and onto the end-to-end distance at 300.9K for cGAGAg and cUUCGg.\\
\textbf{Figure S6} Change in folding free energy upon addition of a local Gaussian penalty to torsion potential terms. \\
\textbf{Figure S7} Potential energy and free energy profiles of torsion angles upon the addition of cosine modifications.\\
\textbf{Figure S8} Free energy surfaces projected onto the eRMSD from native and end-to-end distance upon the addition of cosine modifications.\\
\textbf{Figure S9} Comparison between potential energies calculated using the classical AMBER force field and high-level quantum mechanical calculations.\\
\end{suppinfo}


\bibliography{references}

\end{document}